\documentclass[a4paper]{jpconf}
\usepackage{graphicx}
\newcommand{\be}{\begin{eqnarray}}
\newcommand{\ee}{\end{eqnarray}}

\renewcommand{\H}{\hat H}
\renewcommand{\d}{\mbox{{\rm d}}}
\def\comment#1{}

\begin{document}

\title{Can quantum mechanics be an emergent phenomenon?}

\author{Massimo Blasone$^a$, Petr Jizba$^{b,c}$, Fabio Scardigli$^{d,e}$}
\address{$^a$ INFN, Gruppo Collegato di Salerno; DMI, Universit\`a
di Salerno, Fisciano (SA) - 84084 Italy}
%
%
\address{$^b$ ITP, Freie Universit\"{a}t Berlin, Arnimallee 14
D-14195 Berlin, Germany;}
\address{$^c$ FNSPE, Czech Technical University in Prague,
B\u{r}ehov\'{a} 7, 115 19 Praha 1, Czech Republic}
%
%
\address{$^d$ Leung Center for Cosmology and Particle Astrophysics (LeCosPA),\\
$^{\,\,\,\,\,}$Department of Physics, National Taiwan University, Taipei 106, Taiwan;}
\address{$^e$ Yukawa Institute for Theoretical Physics, Kyoto University,
Kyoto 606-8502, Japan}
\ead{blasone@sa.infn.it}
\ead{jizba@physik.fu-berlin.de}
\ead{fabio@phys.ntu.edu.tw}
\begin{abstract}
We raise the issue whether conventional quantum mechanics, which is not
a hidden variable theory in the usual Jauch-Piron's sense, might nevertheless be
a hidden variable theory in the sense recently conjectured by G.~'t~Hooft in his
pre-quantization scheme. We find that quantum mechanics might indeed have a fully
deterministic underpinning by showing that Born's rule naturally emerges
(i.e., it is not postulated) when 't~Hooft's Hamiltonian for be-ables is combined
with Koopmann--von Neumann operatorial formulation of classical physics.
\end{abstract}
%
%
\section{Introduction\label{Sec.1}}
\vspace{3mm}
Although quantum mechanics (QM) has been the undisputed basis for most of progress
in fundamental physics during the last 90 years or so, the extension of the current
theoretical frontier to Planck's scale physics, and recent enlargements of our experimental
capabilities, may make the 21st century the period in which possible limits of quantum
theory will be subjected to a thorough scrutiny. The basic premise of this paper is that
QM is actually not a complete ontological system, but in fact it represents a very
accurate low-energy approximation to a deeper level of dynamics. But
what exactly the ``deeper level dynamics"  should be? There is
a growing interest in this {\em bone of contention} which is partially fueled
by the belief that in order to make a convincing synthesis of QM and general relativity
(GR) a new conceptual paradigm is needed to tackle physics at very small space-time scales.
There is, however, a fundamental discord in how this should be achieved. One way
of thinking maintains that at high energies the rules of GR should be changed/modified while
the rules of quantum physics should be kept untouched. Such a view is typically justified by
arguing that QM is the most precise theory ever, being accurate to about one part in
$10^{11}$ (for the latest precision tests of QED see, e.g. Ref.~\cite{pdg}). As pointed by
R.~Penrose this view is deceptive~\cite{pen-hawk}: in fact GR has now been experimentally
tested by measuring the orbit-period slowdown in the Hulse-Taylor binary pulsar PSR 1913 + 16
with accuracy to one part in $10^{14}$ --- and this precision is apparently limited only by the
accuracy of clocks on Earth~\cite{taylor}. Another way of thinking holds that it is rather QM that
should be replaced at high energy scales with a more fundamental paradigm, while GR may, or may not,
be modified at such high energies.
Along this line, motivated by the black-hole thermodynamics, G. 't~Hooft has hinted
~\cite{hooft1} that the fundamental rules should be, at very high energy scale --- perhaps at
the Planck scale --- deterministic.

There is a long standing suspicion towards determinism that might underlie QM which
mainly steams from Bell's inequality (BI)~\cite{Bell} and its various generalizations~\cite{Bell2}.
While deterministic theories always satisfy Bell's inequality, QM systems evidently experimentally
violate BI~\cite{Aspect1,Tittel,Rowe}. This is then interpreted as a proof of the non-existence
of a deterministic
underpinning for QM. At this stage it should be stressed that BI's are not a result about QM.
They simply state that {\em usual} deterministic theories obeying the {\em usual} (Kolmogorovian)
probability theory inevitably satisfy certain inequalities between the mean values of certain
(suitably chosen) observed quantities. By ``usual deterministic theories'' one means theories
that are {\em realistic}, i.e. the system has an intrinsic existence independent of observation,
and {\em local}, i.e. sufficiently separated measurements should not influence each other.
Though these assumptions about the underlying deterministic nature of QM are certainly intuitively
plausible, they are too much rooted in our everyday laboratory-scale experience.
Planck's scale ($\sim 10^{19}$ GeV, i.e. $\sim 10^{-35}$m) deterministic dynamics
(whatever this means) may, after all, look non-local to a human observer whose observational
scale is at best few TeV ($\sim10^{-19}$m). In fact, according
to 't~Hooft~\cite{hooft2}, the enormous amount of information that would be lost
in the process of ``coarse graining" from $10^{19}$GeV to $10^3$GeV would lead to
formation of equivalence classes, in the sense that many distinct
states of the Planckian-scale dynamics would go into a single state of the observational-scale dynamics.
For example, in Ref.~\cite{hooft2} is proposed that, when the
dynamical equations at be-ables level describe the configuration-space of a
chaotic system, the equivalence classes could be related
to the stable orbits of such system (e.g., limit cycles). The mechanism responsible
for clustering of trajectories to equivalence classes is identified
by 't Hooft as information loss: after a while one cannot retrace
back the initial conditions of a given trajectory, one can only say
at what attractive trajectory it will end up.
Explicit examples of be-able systems that give rise, at
a macroscopic level, to a genuine quantum behavior can be found in
Refs.~\cite{hooft1,hooft2,blasone2,blasone1}.
Applications of the outlined scenario have been given,
e.g., in Refs.~\cite{blasone2,blasone1,scard,Elze:2005hs,Elze:2008a}.
For other approaches see also Refs.~\cite{Wetterich:2002fy,Adler:2005,Smolin:2006bw}

The passage to quantum theory without conventional quantization procedure
is called by 't~Hooft {\em pre-quantization}.
In what follows we wish to use the pre-quantization concept to tackle the issue of Born's rule
from first principles, i.e. without need to postulate it.
The paper is organized as follows: In Section~\ref{Sec.2} we present some
fundamentals of 't~Hooft pre-quantization method. An extension of this is presented
in the second half of Section~\ref{Sec.2}. There we put forward some ideas concerning
the dependance of 't~Hooft's loss-of-information constraint on observer's energy scale.
A connection with Koopman--von Neumann (KvN) operatorial formulation of classical
physics is discussed in Section~\ref{Sec.3}. There we demonstrate in some detail that
Born's rule is, at the primordial energy scale, merely a convenient mathematical instrument
for doing classical statistical physics, while other, equivalent means are also available.
It is only at low-energy (observational) scales, where the operatorial and deterministic modes
of description substantially depart, that Born's rule starts to have its own independent existence.
Various remarks and speculations are postponed till the concluding Section~\ref{Sec.4}.

\section{Essentials of 't~Hooft's pre-quantization method\label{Sec.2}}
\vspace{3mm}
In this section we briefly outline 't Hooft's continuous-time pre-quantization
method~\cite{hooft2}. A discrete-time version, which employs cellular automata,
is discussed, e.g. in Refs.~\cite{hooft1,blasone1,JizbaIII}, and will not be followed here.
We start with the assumption that the dynamics at the primordial deterministic level is described
by the Hamiltonian
\be
H\ = \ \sum_i p_i f_i({\bf q}) \ + \ g({\bf q})\, . \label{hc}
\ee
Here $f_i$ and $g$ are functions of ${\bf q} = \{q_1,\ldots, q_n\}$.
One should note that the equation of motion for $q_i$, i.e.
\be
\dot{q}_i\ = \ f_i({\bf q})\, , \label{1.2}
\ee
is autonomous, since the $p_i$ variables are decoupled. The above system is obviously
deterministic, however its Hamiltonian is not bounded from below.
It is also worthy of noting that
\begin{eqnarray}
q_i(t + \Delta t) &=& q_i(t) + f_i({\bf q})\Delta t
+ \frac{1}{2}f_k({\bf q} )\frac{\partial^2 H}{\partial p_i \partial q_k}(\Delta t)^2 \ + \ \cdots \,
= \,  F_i({\bf q}(t),\Delta t)\, , \label{1.3}
\end{eqnarray}
where $F_i$ is some function of ${{\bf q}}(t)$ and $\Delta t$ but not ${\bf p}$.
Since (\ref{1.3}) holds for any $\Delta t$ we get the Poisson bracket
\be
\{ q_i(t'), q_k(t)\} = 0 \;\;\;\;\; \mbox{for} \;\;\;\;\;\forall\;\; t,t'\, .\label{1.4}
\ee
Because of the autonomous character of Eq.(\ref{1.2}) one can define a formal Hilbert space
$\mathcal{H}$ spanned by the states $\{|{\bf q}\rangle\}$, and associate with $p_i$ the operator
${\hat p}_i =-i\partial/\partial q_i$. It is not difficult to see that the generator of time
translations (i.e., the ``Hamiltonian" operator), of the form
${\hat H} = \sum_i {\hat p}_i f_i({\hat{\bf q}}) +g({\hat{\bf q}})$
generates precisely the deterministic evolution (\ref{1.2}).
Indeed, we first observe that because ${\hat H}$ is generator of time translations then in
Heisenberg's picture
\be
{\hat q}_i (t + \Delta t) = e^{i\Delta t \hat{H}} {\hat q}_i (t)e^{-i\Delta t \hat{H}}\, , \label{1.5}
\ee
which for infinitesimal $\Delta t$ implies
\be
&&{\hat q}_i (t + \Delta t) -  {\hat q}_i (t)
= i \Delta t[\hat{H}, q_i(t)]\,\;\;\;
\Rightarrow \,\;\;\; \dot{\hat q}_i\,  =f_i(\hat{\bf q})\,.  \label{1.6}
\ee
On the other hand for arbitrary finite $\Delta t$ we have from (\ref{1.5})
\be
{\hat q}_i (t + \Delta t) &=&
\sum_{n=0}^{\infty} \frac{1}{n!}[\hat{H},[\hat{H},[\cdots [\hat{H},{\hat q}_i (t)]]\cdots]] \;
= \;\tilde{F}_i({\hat{\bf q}}(t),\Delta t)\, . \label{1.7}
\ee
In Eq.(\ref{1.7}) $\hat{H}$ appears in the generic term of the sum $n$ times.
On the other hand $\tilde{F}_i$ is some function of ${\hat{\bf q}}(t)$ and $\Delta t$ but not
${\hat{\bf p}}$, which immediately implies that
\be
[{\hat q}_i(t), {\hat q}_j(t')] \ = \ 0\, , \label{1.8}
\ee
for any $t$ and $t'$ (this in turn gives $F_i = \tilde{F}_i$).
Result (\ref{1.8}) shows that the Heisenberg equation of motion for ${\hat q}_i(t)$ in
the ${\bf q}$-representation is identical with the $c$-number dynamical Eq.(\ref{1.2}).
This is because ${\hat q}_i(t+ \Delta t)$ and ${\hat q}_i(t)$ commute, and hence
${\hat q}_i(t+ \Delta t)$,  ${\hat q}_i(t)$, $f_i(\hat{\bf q})$ and also
$\d {\hat{q}}_i(t)/\d t$ can be simultaneously diagonalized.
In this diagonal basis we get back the $c$-numbered autonomous Eq.(\ref{1.2}).
In other words, operators ${\hat q}_i$ evolve deterministically even after ``quantization''.
This evolution is only between base vectors. From the Schr\"{o}dinger-picture point of view
this means that the {\em state vector} evolves smoothly from one {\em base vector} to another
(in Schr\"{o}dinger picture base vectors are time independent and fixed).
So at each instant the state vector coincides with some specific base vector.
Because of this, there is {\em no} non-trivial linear superposition of the state vector
in terms of base vectors and hence {\em no} interference phenomenon shows up when measurement
of ${\bf q}$-variable is performed. Dynamical variables fulfilling Eq.(\ref{1.8}) were first
considered by Bell~\cite{Bell} who called them \emph{be-ables} as opposed to observed
dynamical variables which are in QM called \emph{observables}.

\begin{figure}[t]
%
\centerline{\includegraphics[width=26pc]{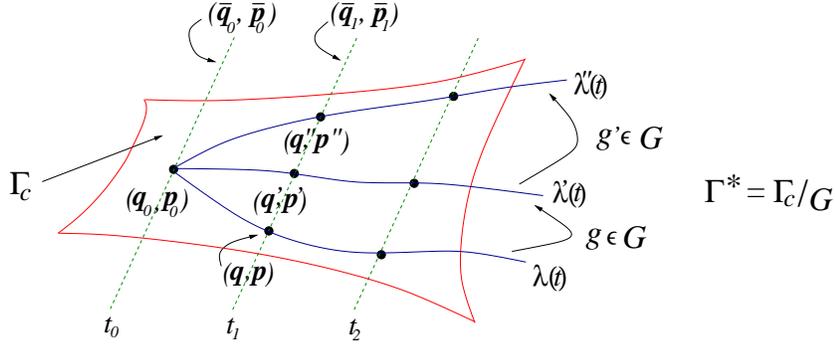}}
%
\caption[]{\label{gauge}\small{Pictorial representation of the gauge
freedom inherent in the evolution $\d q_i/ \d t = \{q_i, H_e\}$.
Here $({\bf q},{\bf p})$, $({\bf q}',{\bf p}')$, $({\bf q}",{\bf p}")$, $\ldots$
describe the same physical state $(\bar{\bf q}_1,\bar{\bf p}_1)$.
$g$ are elements of the gauge group generated by $\Phi_E$.
For simplicity's sake we do not explicitly write down the sub-index $E$.}}
\vspace{0.2cm} \hrule
%
\end{figure}

Let us now come back to Eq.(\ref{hc}).
One may immediately notice that $H$ is unbounded from below. This fact should not disturb us
too much since actual dynamics of be-ables is described by Eq.(\ref{1.2}).
The $H$ was merely introduced to set up the parallel operatorial formulation of the be-able dynamics.
In the following section we will see that this Hamiltonian has yet another useful role, namely it
helps to formulate a classical statistical mechanics for be-ables.
On the observational (i.e. emergent) level the Hamiltonians are key objects providing both the equations
of motions and the energy of the system. The {\em emergent} Hamiltonians must be bounded from below.
The concept of lower bound is, in 't~Hooft's proposal, just an emergent property formed during the coarse
graining of the be-able degrees of freedom down to the observational ones.
We can devise a simple toy model mechanism showing how a lower bound for $H$ may develop.
To this end consider  $\rho({\hat q})$ to be some positive function of ${\hat q}_i$
(but not ${\hat {\bf p}}$) with $[\hat{\rho}, \hat {H} ] = 0$.
One then defines the splitting
\be
&&\mbox{\hspace{-9mm}}\H \ = \ \H_+ - \H_-, \nonumber \\
&&\mbox{\hspace{-9mm}}\H_+ \ = \ \left(\hat{\rho} + \H \right)^2\frac{\hat{\rho}^{-1}}{4},\quad \H_- \
= \ \left(\hat{\rho} - \H\right)^2 \frac{\hat{\rho}^{-1}}{4}, \label{2.5}
\ee
where $\H_+$ and $\H_-$ are positive-definite operators satisfying
\be
[\H_+,\H_-] \ = \ [\hat{\rho},\H_{\pm}] \ = \ 0 . \label{comm}
\ee
At this stage we introduce the ``coarse-graining" operator $\hat{\Phi}$ that describes the
loss of information occurring during the passage from the be-able to observational scale.
One possible choice is~\footnote{This form might be inspired by the entropy
considerations reported in ~\cite{blasone2}. We shall discuss this issue elsewhere.}
\be
\hat{\Phi}_E \ = \ (1- e^{-(E_p - E)/E})\hat{H}_-\, ,\label{1.12}
\ee
where $E$ refers to the observer's energy scale, while $E_p$ is the be-able energy scale,
which we take to be Planck's scale. The operator $\hat{\Phi}_E$ is then implemented as a
constraint on the Hilbert space $\mathcal{H}$. So at each observational energy scale $E$,
the observed physical states $|\psi\rangle_{phys}$ are given by the condition
\be
\hat{\Phi}_E |\psi\rangle_{phys}\ = \ 0 \, . \label{1.13}
\ee
This equation identifies the states that are not affected by the coarse graining, i.e.,
states that are still distinguishable at the observational scale $E$.
The $c$-number factor $(1- e^{-(E_p -E)/E})$ is $not$ irrelevant and one cannot use
directly $\hat{H}_-$ instead of $\hat{\Phi}_E$. In fact the constraint (\ref{1.13}) is,
according to Dirac's classification of constraints~\cite{sunder}, a first class primary
constraint because $[\hat{\Phi}_E,\hat{\Phi}_E]= 0$ and $[\hat{\Phi}_E,\hat{H}] = 0$.
First-class constraints generate gauge transformation and hence not only restrict the full
Hilbert space (from $\mathcal{H}$ to $\mathcal{H}_c$) but also produce equivalence classes of states,
which are in general non local (e.g., points with a space-like separation are identified).
Two states belong to the same class if they can be transformed into each other by a gauge
transformation with the generator $\hat{\Phi}_E$.
Let $\mathcal{G}_E$ be the one parameter group of these gauge transformations generated
by $\hat{\Phi}_E$. The equivalence classes built out of such gauge transformations
represent at each scale $E$ the physical states (i.e., observables).
If $\mathcal{O}_E$ denotes the space spanned by the observables, we can identify
$\mathcal{O}_E$ with a quotient space
\be
\mathcal{O}_E \ = \ \mathcal{H}_c/\mathcal{G}_E\, . \label{2.4}
\ee
The quotient space $\mathcal{O}_E$ (its structure and dimensionality) depends on the
energy scale $E$. In particular at the level of be-ables where $E = E_p$ the constraint
$\hat{\Phi}_E$ is identically zero and the space of observables is directly the
Hilbert space $\mathcal{H}$. On the other hand, when $E \ll E_p$, e.g. at scales available
to a human observer, one has $\hat{\Phi}_E= \hat{H}_-$. The latter is the constraint
originally considered by 't~Hooft~\cite{hooft1,hooft2}. In such a case the resulting physical
state space, i.e. the space of observables, has the energy eigenvalues that are trivially
bounded from below owing to
\be
\H|\psi\rangle_{phys} \ = \ \H_+|\psi\rangle_{phys} \ = \ \hat{\rho}|\psi\rangle_{phys}\, .
\ee
Thus, in the Schr{\"o}dinger picture the equation of motion
\be
\frac{\d}{\d t}|\psi_t\rangle_{phys}\ = \ -i\H_+|\psi_t\rangle_{phys}\, , \label{2.6}
\ee
has only positive frequencies on physical states.
Above coarse-graining procedure was so far implemented in an operatorial mode of description
of the be-able dynamics. As a result we have obtained on observational energy scales
a Schr\"{o}dinger equation with a ``well behaved" (i.e., bounded from below) emergent Hamiltonian.
Constraining procedure can be, in parallel, also applied to a deterministic mode of description
of the be-able dynamics. This can be done through the Dirac--Bergmann algorithm~\cite{Dir2}.
Here one implements the constraint $\Phi_E$ directly on the phase space $\Gamma$ spanned by
${\bf p}$ and ${\bf q}$. This results in the constrained hyper surface $\Gamma_c$.
Because $\Phi_E$ Poisson-commutes with itself and with $H$, it is a first class primary
constraint~\cite{Dir2}. Similarly as in the operatorial case, this means that $\Phi_E$ not
only restricts the phase space but also provides classes of equivalence for possible evolutions.
This is because $\Phi_E$ is unable to determine the Lagrange multiplier in the ``extended"
Hamiltonian $H_e({\bf q}, {\bf p}) = H({\bf q}, {\bf p})+ \lambda(t) \Phi_E({\bf q})$.
In particular, the self-consistency equation
\be
0 \ = \ \frac{\d \Phi_E}{\d t} \ = \ \{\Phi_E, H_e\} \
= \ \{\Phi_E,H\} \ + \ \lambda(t) \{\Phi_E, \Phi_E\}\, ,
\ee
is trivially fulfilled without giving any information on $\lambda(t)$.
This leads to the class of equivalent dynamics described by the equation
$\d q_i/\d t = \{q_i, H_e\}$. Each of the possible trajectories in the class is
parameterized by different $\lambda(t)$, see Fig.\ref{gauge}.
Two points belong to the same class if they can be transformed into each other by
gauge transformation generated by $\Phi_E$. For infinitesimal transformations
this means that for two equivalent dynamics
with slightly different $\lambda(t)$
\be
\delta {\bf{q}}(t) = {\bf{q}}_{\lambda(t)} - {\bf{q}}_{\lambda'(t)} = \varepsilon \{{\bf{q}}(t), \Phi_E \}\, ,
\ee
with $\varepsilon = t(\lambda - \lambda')$.
Gauge classes can be identified by choosing a single representative element from each class.
This is done by a gauge-fixing condition $\chi$ such that $\{\Phi_E, \chi\}\neq 0$.
The emergent physical space (space of classically observed degrees of freedom at scale $E$)
is a quotient space $\Gamma^*_E =\Gamma_c/\mathcal{G}_E$. It should be noted that the identification
of the equivalent classes is a non-local procedure, since at a given instant one generally identifies
also points with a space-like separation. This is, in spirit, deeply different from the
quotioning procedure used in gauge field theories. There the quotient space is obtained from
the configuration space that in itself is non-physical (it contains huge amount of physically
equivalent configurations). The physical space emerges only after one identifies each equivalence
class with a distinct physical configuration. In contrast, in our case we start with the
configuration space (space of be-able dynamics) that is {\em a priori physical} and the
identification of equivalent classes is done only because we wish to find an ``effective"
(or emergent) description at some energy scale $E$ that is much lower than the be-ables scale.
%
%
\section{Koopman--von Neumann formulation and 't~Hooft's be-able dynamics\label{Sec.3}}
\vspace{3mm}
Let us come back to 't~Hooft's Hamiltonian (\ref{hc}) and try to understand its role a bit more.
There are three immediate issues that should be addressed. These are: the issue of hermiticity,
the issue of the canonical momenta $p_i$, and finally the issue of the ensuing state-vectors
and related Born's rule. The hermiticity is not a real problem at the energy scale $E_P$.
There the dynamics is driven by Eq.(\ref{1.2}) and the Hamiltonian is only a formal tool that
allows to generate the corresponding evolution equations. Should one wish to have hermitian Hamiltonian
from the very scratch, one may compensate for the non-hermitian ordering between
$\hat{p}_i$ and $f_i(\hat{\bf q})$ by adding the (non dynamical) function $g(\hat{\bf q})$
to $\hat{H}$. This should be done in such a way that
$g^\dag(\hat{\bf q}) - g(\hat{\bf q}) =\sum_i[\hat{p}_i, f_i(\hat{\bf q})]$,
so that $\hat{H}^\dag =\hat{H}$. In the particular case when $g(\hat{\bf q})$ is anti-hermitian
the Hamiltonian is not only hermitian but also Weyl ordered. Indeed, in this case we have
\begin{eqnarray}
\hat{H} \ = \ \sum_i \hat{p}_i f_i(\hat{\bf q}) +g(\hat{\bf q})
&=&\mbox{$\frac{1}{2}$}\sum_i \left(\hat{p}_i f_i(\hat{\bf q}) +
f_i(\hat{\bf q})\hat{p}_i\right) +
\mbox{$\frac{1}{2}$}\sum_i[\hat{p}_i,f_i(\hat{\bf q})] + g(\hat{\bf q})\nonumber \\
&=& \mbox{$\frac{1}{2}$}\sum_i \left(\hat{p}_i f_i(\hat{\bf q}) +
f_i(\hat{\bf q})\hat{p}_i\right) = W\!\left(\sum_i \hat{p}_if_i(\hat{\bf q}) \right)\!\!.\label{Vb1}
\end{eqnarray}
In this connection it is interesting to notice that the Weyl form (\ref{Vb1}) also enters in the
Pontryagin's approach~\cite{pontryagin62} to quantization of non-Hamiltonian systems described
by equations (\ref{1.2}). The hermiticity itself is, however, an important concept at low energies,
where the emergent Hamiltonian is directly interpreted as the energy of the system.
However, in the existent models of 't~Hooft's constraining conditions the
emergent Hamiltonian is hermitian. The canonical momentum $p_i$ is at the be-able level just a dummy
variable which merely allows to set up a Hamiltonian formalism. In the previous section we have
seen that it also allows to formulate an operatorial version of the be-able dynamics. It should be noticed
that, while at the Planck energy $E_P$ the Poisson bracket (or alternatively the commutator)
of the be-able position
with its {\em kinetic} momentum vanishes,
i.e. $\{q_i(t), p^{kin}_j(t)\} \propto \{q_i(t),\dot{q}_j(t)\} = 0 $, on the contrary
the corresponding Poisson bracket with canonical momentum  is the canonical one, $\{q_i(t), p_j(t)\} = 1$.
Since the be-able dynamics (\ref{1.2}) is formulated on the tangent rather than cotangent bundle,
it is the kinetic  rather than canonical momenta that is physically a more relevant object at $E_P$.
The third and more pressing issue is the Born rule. We now show that the Born rule is  closely related
to Koopman--von Neumann's (KvN) operatorial formulation of classical physics~\cite{KvN}~\cite{Elze}.
To this end we can ask ourselves
how an observer ``living'' in the be-able world would do statistical physics on systems described
by Eq.(\ref{1.2}). In fact, the simplest way is to go through the Hamiltonian (\ref{hc}).
In this case Koopman and  von Neumann found a simple recipe for defining the
probability density function. Following KvN we define a ``wave function'' $\psi({\bf p}, {\bf q},t)$
that evolves in time with the Liouville operator, i.e., the Hamilton field (summation over $i$ is understood)
\be
\hat{\mathcal{H}} \ = \ - i \partial_{p_i}H({\bf p}, {\bf q})\partial_{q_i} +
i \partial_{q_i} H({\bf p}, {\bf q}) \partial_{p_i}\, ,
\ee
according  to the equation
\be
i \frac{\partial }{\partial t }\psi \ = \ \hat{\mathcal{H}}\psi\, . \label{3.1}
\ee
The vectors $\psi$ are complex wave functions on the phase space $\Gamma = ({\bf p}, {\bf q})$,
with the normalization $\int \d {\bf p}\d {\bf q} \ \!|\psi({\bf p}, {\bf q})|^2 =1$.
The corresponding complex conjugation turns Eq.(\ref{3.1}) to
\be
i \frac{\partial }{\partial t}\psi^* \ = \ \hat{\mathcal{H}}\psi^*\, . \label{3.2}
\ee
By multiplying (\ref{3.1}) with $\psi^*$ and (\ref{3.2}) with $\psi$, and adding them together we obtain
\begin{eqnarray}
i \frac{\partial}{\partial t} \varrho\
= \ \hat{\mathcal{H}}\varrho \;\;\;\;
\Leftrightarrow \;\;\;\; \frac{\partial}{\partial t} \varrho \
= \ (-\partial_{p_i}H \partial_{q_i} +
\partial_{q_i} H \partial_{p_i})\varrho\, , \label{3.3}
\end{eqnarray}
where we have defined $\varrho({\bf p}, {\bf q}) = \psi^*({\bf p},{\bf q})\psi({\bf p}, {\bf q})$.
The second equation is nothing but the well-known Liouville equation for the probability density function
of a (classical) statistical system whose dynamics is driven by a (classical) Hamiltonian $H$.
If one defines the scalar product between two wave functions $\psi_1$ and $\psi_2$ as
$\langle\psi_1|\psi_2\rangle = \int\d {\bf p}\d {\bf q} \ \! \psi_1^*\psi_2$,
then it is easy to show that
$\langle \psi_1 |\hat{\mathcal{H}} \psi_2\rangle = \langle\hat{\mathcal{H}} \psi_1 |\psi_2\rangle$,
i.e., the Liouvillian $\hat{\mathcal{H}}$ is a self-adjoint operator, which implies that
the norm of any state is conserved during the evolution.
The latter is consistent with the interpretation of
$\varrho =\psi^*\psi$ as a probability density function on the phase space $\Gamma$.
Now, since the ${\bf p}$ variables are in 't~Hooft's proposal only dummy variables
(true degrees of freedom are be-ables, i.e., variables ${\bf q}$) the relevant density matrix
is the reduced density matrix $\tilde{\varrho}({\bf q}) = \int \d {\bf p} \ \! \varrho({\bf p},{\bf q})$.
By using 't~Hooft's Hamiltonian (\ref{hc}) and integrating (\ref{3.3}) over ${\bf p}$
we obtain
\be
\frac{\partial}{\partial t} \tilde{\varrho}({\bf q})
&=& - f_i({\bf q})\partial_{q_i} \tilde{\varrho}({\bf q}) -
[\partial_{q_i} f_i({\bf q})]\tilde{\varrho}({\bf q}) \ =
\ - \partial_{q_i}(f_i({\bf q})\tilde{\varrho}({\bf q}))\nonumber \\
&=& -i \hat{p}_i f_i({\bf q}) \tilde{\varrho}({\bf q}) \ = \ -i
\hat{H}({\bf p}, {\bf q}) \tilde{\varrho}({\bf q})\,  , \label{3.4}
\ee
where $\hat{H}$ is 't Hooft's Hamiltonian without the $g$ term. If now we define
\be
\tilde{\psi}({\bf q}) =  \int_{-\infty}^{\infty} \d {\bf p} \ \!\psi({\bf p}, {\bf q})\, , \label{3.5}
\ee
we can use Eq.(\ref{3.1}) (respectively Eq.(\ref{3.2})) to compute the evolution of
$\tilde{\psi}({\bf q})$ (respectively $\tilde{\psi}^*({\bf q})$) under the 't~Hooft's Hamiltonian
$\hat{H}$, so that we obtain
%
\be
i \frac{\partial}{\partial t} \tilde{\psi}({\bf q}) \ =
\ \hat{H}({\bf p}, {\bf q})\tilde{\psi}({\bf q})\;\;\;\;\;\;\;
\mbox{and} \;\;\;\;\;\;\; i\frac{\partial}{\partial t} \tilde{\psi}^*({\bf q}) \  =
\ \hat{H}({\bf p},{\bf q})\tilde{\psi}^*({\bf q})\, . \label{3.6}
\ee
Multiplying the first by $\tilde{\psi}^*({\bf q})$ and the second by $\tilde{\psi}({\bf q})$,
 summing together, and taking advantage of the particular form of 't Hooft's Hamiltonian, we get
\be
\frac{\partial}{\partial t} (\tilde{\psi}^*({\bf q})\tilde{\psi}({\bf q}))
= \ -i\hat{H}({\bf p},{\bf q})(\tilde{\psi}^*({\bf q})\tilde{\psi}({\bf q}))
\ee
which, confronted with (\ref{3.4}), bring us to the identification
$\tilde{\varrho}({\bf q}) = \tilde{\psi}^*({\bf q})\tilde{\psi}({\bf q})$

This is precisely the (configuration-space) Born rule.
While in the previous section we have seen
that the be-able dynamics can be equivalently described both by classical or (quantum-like)
operatorial means, this section reinforces the picture by showing that this parallelism can be
extended also to statistical physics. The parallel mode of description of course breaks
when constraint (\ref{1.13}) is imposed.
We thus see that 't~Hooft's Hamiltonian has a privileged role among
classical Hamiltonians. In fact, as shown in Ref.~\cite{Jizba}, there are no other
Hamiltonian systems with the peculiar property that their full quantum evolution coincides
with the classical one.

We should finally note that the above canonical scenario has also a path-integral counterpart.
In this case the KvN operatorial version of the deterministic system (\ref{hc}) is taken over
by the classical path integral of Gozzi {\em et al}~\cite{Jizba,Gozzi}.
In fact, because both $\tilde{\psi}({\bf q})$ and $\tilde{\varrho}({\bf q}) =|\tilde{\psi}({\bf q})|^2$
satisfy the same
equation of the motion -- Schr\"{o}dinger's equation, the solutions can be written through the
same evolution kernel, i.e.
\be
\tilde{\varrho}({\bf q},t) = \int \d {\bf q}' K({\bf q},t; {\bf q}',t')\tilde{\varrho}({\bf q}',t')\, ,
\quad {\rm and} \quad
\tilde{\psi}({\bf q},t) = \int \d {\bf q}' K({\bf q},t; {\bf q}', t')\tilde{\psi}({\bf q}',t')\, . \label{3.7b}
\ee
Although both above kernels represent formally
identical expressions, they are qualitatively different. The first kernel in (\ref{3.7b}) corresponds to
a transition {\em probability}, while the second corresponds to a transition {\em amplitude}.
The fact that the same kernel propagates both $\tilde{\psi}$ and $|\tilde{\psi}|^2$ was, in the KvN framework, firstly
recognized and discussed in Refs.~\cite{Gozzi}. The corresponding discussion for 't~Hooft
Hamiltonian system was done in Ref.~\cite{Jizba}. The evolution kernel itself has the path integral
representation
\be
K({\bf q},t; {\bf q}', t')
= {\mathcal{N}} \int_{{\bf{\xi}}(t) =
{\bf{\xi}}_1}^{{\bf{\xi}}(t') = {\bf{\xi}}_2}\!
{\mathcal{D}}{\bf{\xi}} \ \exp \left[i\! \int_{t}^{t'}\! \d
t~L({\bf{\xi}},
\dot{\bf{\xi}})\right] = {{\mathcal{N}}} \int_{{\bf{q}}(t) =
{\bf{q}}}^{{\bf{q}}(t') = {\bf{q}}'} {\mathcal{D}}{\bf{q}} ~\prod_i
\delta[ \dot{q}_i-f_i({\bf{q}})]\, .
\label{3.8}
\ee
Here we have defined $2N$ configuration-space coordinates as
\begin{eqnarray}
\xi_i ~= ~p_i, \;\;\; i = 1, \ldots, N \quad\quad {\rm and} \quad\quad
\xi_i ~= ~q_i, \;\;\; i = N+1, \ldots, 2N\, ,
\end{eqnarray}
and the Lagrangian
\begin{eqnarray}
L({\bf{\xi}}, \dot{\bf{\xi}}) \ =
\ \mbox{$\frac{1}{2}$} \xi^i\omega_{ij} \dot{\xi}^j - H({\bf{\xi}})\, , \label{3.26}
\end{eqnarray}
(${\bf{\omega}}$ is the $2N\times 2N$ symplectic matrix). The factor ${\mathcal{N}}$ is a
normalization factor, and the measure can be explicitly written as
\begin{equation}
{\mathcal{N}} \int_{{\bf{\xi}}(t) =
{\bf{\xi}}_1}^{{\bf{\xi}}(t') ={\bf{\xi}}_2} {\mathcal{D}}{\bf{\xi}} \
= \ {{\mathcal{N}}}\int_{{\bf{q}}(t) = {\bf{q}}_1}^{{\bf{q}}(t')
= {\bf{q}}_2}{\mathcal{D}}{\bf{q}} {\mathcal{D}}{{\bf{p}}} \, .\label{@}
\end{equation}
The kernel representation (\ref{3.8}) corresponds to the classical path integral
of Gozzi {\em et al}~\cite{Gozzi}. The constraining procedure can be now applied directly
to the path integral (\ref{3.8}) either through Dirac--Bergmann~\cite{Dir2} or Faddeev--Jackiw
procedure~\cite{Jizba,F-J,JizbaII}. The actual constraining mechanism is however different when
applied to path integrals representing transition amplitudes or to path integrals representing
transition probabilities. In particular, for transition probabilities one should rewrite
the path integral in the Feynman--Vernon closed-time-path form~\cite{FV}, and then apply the
constraint to each time branch independently. This lead to a different emergent kernel than
in the case of transition amplitudes. So, similarly as in the operatorial framework, in
the path-integral formulation the constraint (\ref{1.13}) removes the dichotomy of classical-quantal
formulation of the
be-able dynamics. It should be stressed that the symplectic structure in both kernels will be,
however, still preserved, yet now with an effective Hamiltonian and with different degrees of
freedom~\cite{JizbaII}.

%
\section{Conclusions and outlooks\label{Sec.4}}
\vspace{3mm}
In this paper we have demonstrated that when KvN operatorial formulation of classical
physics is combined with 't~Hooft's Hamiltonian for be-ables, one naturally arrives at
the Born's rule without postulating it. This result substantiates 't~Hooft's pre-quantization scheme
and affirmatively answer to the question whether QM can be a hidden variable theory.
It should be stressed that the be-able dynamics is fully local (see Eq.(1)).
The non-locality that provides the loophole in Bell inequalities is realized in 't~Hooft's
pre-quantization by identifying the physical space of observables $\Gamma_E^*$ with a quotient
space ${\Gamma_c}/{\mathcal{G}}_E$ that has a non-local structure with respect to the physical space
of be-ables $\Gamma$.

The picture one can draw from our discussion
\begin{figure}[t]
%
\centerline{\includegraphics[width=20pc]{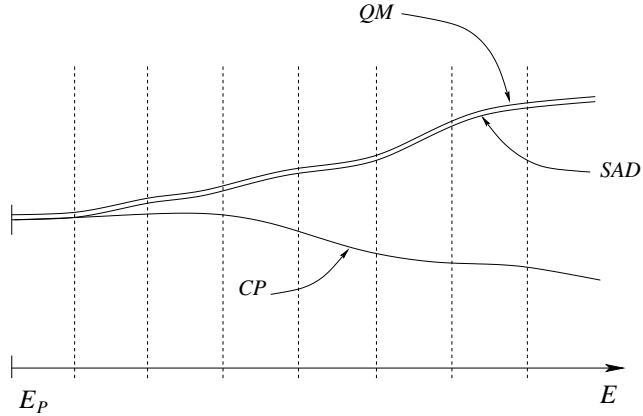}}
\caption[]{\label{div}\small{Schematic representation of the emergent quantum description.
At the primordial level $E_P$ both classical and quantal descriptions are equivalent.
Also, both the state $\psi$ and the deterministic density matrix $\varrho$ are described
via the same evolution kernel $K$. As the energy scale lowers, the information is continuously
lost in the process of coarse graining and the observer sees only effective
dynamics, i.e. equivalence classes of primordial dynamics. The dashed lines remind that
constraint (\ref{1.13}) must be applied at all energy scales.
Operatorial and deterministic description are coarse-grained
together and at low energies are perceived as QM and SAD, respectively.
On the statistical level SAD is described via a density matrix with a symplectic evolution kernel.
The latter is what the observer perceives as CP.}}
\vspace{0.2cm} \hrule
%
%
\end{figure}
is the following (see Fig.\ref{div}): at the primordial scale $E_P$ the dynamics is purely deterministic
and it can be equivalently tackled with classical (description on the level of single trajectories)
and operatorial (description on the level of evolution in the Hilbert space of states) means.
As one lowers the energies, the typical length and time scales are getting larger, and the corresponding
description is getting coarse-grained. This leads to a huge information loss and
brings one to an effective level of description.
A typical observer at $E\ll E_P$ has then at his/her disposal the operatorial mode of
description --- the Schr\"{o}dinger equation (\ref{2.6}) with the effective
Hamiltonian $\hat{H}_+$ , which he/she calls {\em Quantum Mechanics}.
On a parallel footing, there is also available  to him/her a non-operatorial, non-local, coarse-grained
description of the deterministic system that propagates at energy $E_P$. This non-local
imprint of the be-able dynamics\footnote{We denote the non-local
imprint of the be-able dynamics as SAD to stress its formal connection with Einstein's
``spooky action at distance''.} (SAD) has,
however, the statistical density matrix that is described with the standard classical
Liouville equation, or equivalently with a classical kernel for transition probabilities.
The kernel itself can be represented via path integral with the Lagrangian that exhibits an explicit
symplectic structure. It is thus natural for our observer to define
as {\em Classical Physics} (CP) the physics experienced at the macroscopic {\em statistical} level.
This emergent macroscopic description exhibits a symplectic structure which is then identified as a typical
signature of CP.
In this view, one should
not be surprised that BI's are violated at an observational scale. This is because our
understanding of CP (on which Bell's inequalities are ultimately based) is not derived from
SAD dynamics alone (where BI's could be easily violated) but only from its statistical behavior.

\ack
We are particularly grateful to G.~'t~Hooft and H.~Kleinert for inspiring conversations.\\
M.B.acknowledges partial financial support from MIUR and  INFN.
F.S. thanks Japan Society for the Promotion of Science for support under the fellowship P06782,
National Taiwan University and LeCosPA for support under the research contract 097510/52313,
and ITP Freie Universit\"{a}t Berlin for warm hospitality.
P.J. would like to acknowledge financial supports from the Ministry of Education
of the Czech Republic (research plan no. MSM 6840770039) and from the Deutsche Forschungs Gemeinschaft
(grant Kl256/47).

\end{document}